\documentclass[apj]{emulateapj}

\def\lesssim{\mathrel{\hbox{\rlap{\hbox{\lower4pt\hbox{$\sim$}}}\hbox{$<$}}}}
\def\gtrsim{\mathrel{\hbox{\rlap{\hbox{\lower4pt\hbox{$\sim$}}}\hbox{$>$}}}}
\def\citeapos#1{\citeauthor{#1}'s (\citeyear{#1})}

\usepackage{graphicx}

\slugcomment{}
\shorttitle{Excitation of Trapped Waves}
\shortauthors{Henisey et al.}

\begin{document}

\title{
Excitation of Trapped Waves in Simulations of Tilted Black Hole Accretion Disks with
Magnetorotational Turbulence
}

\author{Ken B. Henisey and Omer M. Blaes}
\affil{Department of Physics, University of California, Santa Barbara, CA 93106, USA}

\author{P. Chris Fragile}
\affil{Department of Physics and Astronomy, College of Charleston, Charleston, SC 29424, USA}

\and

\author{B\'arbara T. Ferreira}
\affil{Department of Applied Mathematics and Theoretical Physics, University
of Cambridge, Wilberforce Road, Cambridge CB3 0WA, United Kingdom}

\begin{abstract}

We analyze the time dependence of fluid variables in general relativistic, magnetohydrodynamic
simulations of accretion flows onto a black hole with dimensionless spin parameter $a/M=0.9$. We
consider both the case where the angular momentum of the accretion material is aligned with the
black hole spin axis (an untilted flow) and where it is misaligned by $15^\circ$ (a tilted flow).
In comparison to the untilted simulation, the tilted simulation exhibits a clear excess of inertial
variability, that is, variability at frequencies below the local radial epicyclic frequency. We
further study the radial structure of this inertial-like power by focusing on a radially extended
band at $118(M/10M_\odot)^{-1}{\rm Hz}$ found in each of the three analyzed fluid variables.
The three dimensional density structure at this frequency suggests that the power is a
composite oscillation whose dominant components are an over dense clump corotating with the
background flow, a low order inertial wave, and a low order inertial-acoustic wave. Our results
provide preliminary confirmation of earlier suggestions that disk tilt can be an important
excitation mechanism for inertial waves.

\end{abstract}


\section{Introduction}

Quasi-periodic oscillations (QPOs) are observed in the X-ray light
curves of many black hole X-ray binaries (see \citealt{rem06}
for a recent review).  They have also been observed in extragalactic
ultraluminous X-ray sources \citep{str03,str07} and in one active
galactic nucleus \citep{gie08}.  The origin of these
phenomena is still far from clear.  One class of models centers
on trapped wave modes within the accretion flow.  In particular, a
variety of modes have been proposed in hydrodynamic models of geometrically
thin accretion disks  (e.g. \citealt{wag99,kat01}). Axisymmetric
inertial-acoustic modes (``$f$-'' or ``inner $p$-modes'') and axisymmetric
inertial modes (``$g$-'' or ``$r$-modes'')
\footnote{Throughout this paper we will use the term inertial mode and $r$-mode
interchangeably.  We prefer to avoid the use of the term $g$-mode, whose primary
restoring force is generally due to entropy gradients.  In disks, the primary
restoring force for inertial modes is due to specific angular momentum gradients.}
can be trapped due to the existence of the innermost
stable circular orbit around a black hole, which produces a maximum in
the radial profile of the radial epicyclic frequency.  Non-axisymmetric
inertial modes can also be radially trapped between inner and outer
Lindblad resonances.

The very existence of inertial modes in accretion disks has been challenged
recently, as numerical simulations appear to show that magnetorotational
(MRI) turbulence suppresses them.  This has been demonstrated in local shearing
box simulations \citep{arr06}, whose boundary conditions would artificially
trap a discrete axisymmetric mode spectrum.  Power spectra
from these simulations show discrete acoustic modes and a radial epicyclic
oscillation, but no inertial modes.
This behavior has also been seen in global simulations of accretion
disks in a pseudo-Newtonian potential: hydrodynamic disks exhibit
trapped axisymmetric inertial modes, whereas MHD turbulent disks do not
\citep{rey08}.
It appears that axisymmetric inertial modes, which necessarily
have frequencies at or
below the local radial epicyclic frequency, are particularly vulnerable
to nonlinear damping by MRI turbulence, which has a power spectrum that
peaks near the orbital frequency.  Non-axisymmetric inertial modes
can have higher frequencies and might be less vulnerable to MRI turbulence,
but there has as yet been no convincing demonstration of the existence
of a trapped inertial mode spectrum in any MHD simulation.

Another reason why discrete inertial modes may not be present in accretion
disks with MRI turbulence is that even subthermal magnetic fields can extend
the inner trapping radius of the propagation zone of both axisymmetric and
non-axisymmetric inertial modes down to the innermost stable circular orbit (ISCO)
\citep{fu09}.  Unless inertial waves can reflect off the plunging region
of the flow, standing waves will no longer be sustainable.

On the other hand, it has been suggested that warps and eccentricity in disks may play a
fundamental role in a nonlinear excitation mechanism of inertial modes in geometrically thin
accretion disks \citep{kat04a,kat08,fer08}. These large scale deformations interact with trapped
$r$-modes giving rise to intermediate modes. These intermediate modes then couple back to
the warp or eccentricity to produce positive feedback on the original $r$-mode oscillations (see
\citealt{fer08} for a more detailed explanation of this coupling mechanism).

\cite{fra07} recently completed fully general relativistic, MHD simulations of accretion disks
with misaligned black hole spin and disk angular momentum vectors (``tilted disks''). These tilted
disks are strongly warped near the black hole and exhibit
global epicyclic oscillations superimposed on the turbulence in the flow \citep{fra08}. These
oscillations manifest themselves as eccentric orbits of fluid elements in the disk, albeit with a
$180^\circ$ flip in orientation of the elliptical orbits across the midplane of the disk. Tilted
disks may therefore provide favorable conditions for the nonlinear excitation of inertial modes.

Here we report the results of a search for trapped modes in simulations of both an untilted and
tilted disk. In agreement with previous work, we do not observe the presence of modes in the
untilted simulation. However, at the same numerical resolution, we observe significant excess power
with frequencies characteristic of inertial waves in the tilted disk. Because of the complexity of
the background flow, the physical nature of this power is difficult to determine. Nevertheless, its
spatial structure appears to confirm that this is partly inertial in character. This may represent
preliminary confirmation of the warp/eccentricity excitation mechanism, showing that it can be
strong enough to overcome damping due to MRI turbulence. It also shows that trapping, in the sense
of radial localization of power, can be maintained even in the presence of magnetic fields.

This paper is organized in the following manner. In section 2 we briefly review the simulation
parameters and discuss our power spectrum analysis of the simulation data. Section 3 presents power
spectra from both untilted and tilted geometries, evidence that the latter may be dominated by
inertial-like variability, and an analysis of the three dimensional structure of the power at a
particular frequency within this inertial regime. We summarize our conclusions in section 4.

\section{Simulations and Time Series Analysis}

We analyze the two high resolution simulations of \cite{fra07} and \cite{fra08}: an untilted
simulation (90h) with the initial fluid angular momentum aligned with the spin axis of the Kerr
black hole, and a simulation (915h) with the initial fluid angular momentum tilted by $15^\circ$
relative to the spin axis of the hole.  The black hole spacetime in each simulation is described
using Kerr-Schild coordinates, and has a spin parameter $a/M=0.9$.  Apart from the tilt, both
simulations start with identical initial conditions: a torus with pressure maximum centered at
$25R_{\rm G}$, where $R_{\rm G}\equiv GM/c^2$, seeded with weak poloidal magnetic field loops that
follow the equipressure surfaces. Both simulations evolve the disk through ten test particle
orbital periods calculated at the initial pressure maximum (a time unit hereafter generically
referred to as an ``orbit'').

\begin{figure*}
\centering
\epsscale{1.0}
\includegraphics[width=0.48\textwidth]{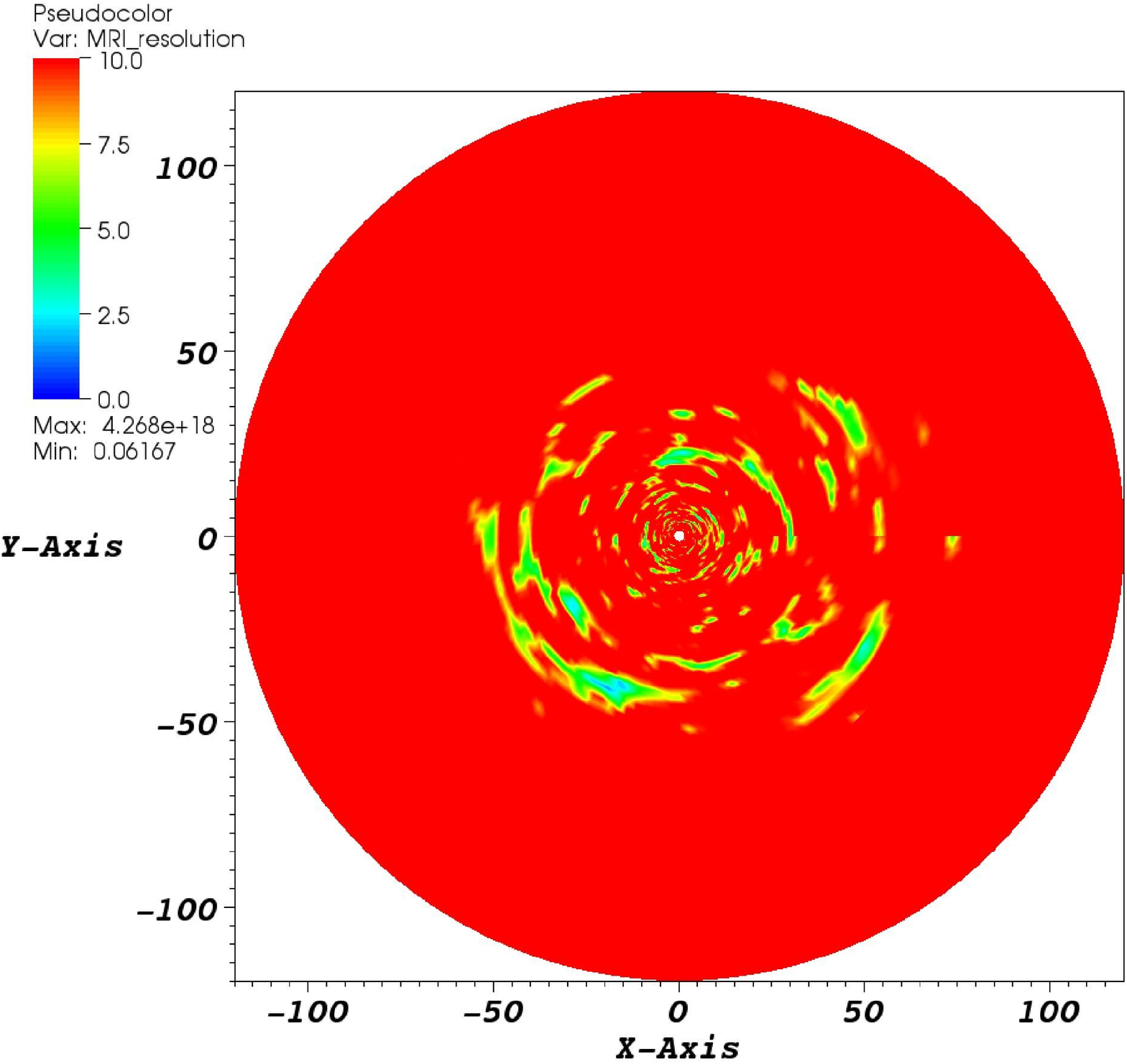}
\includegraphics[width=0.48\textwidth]{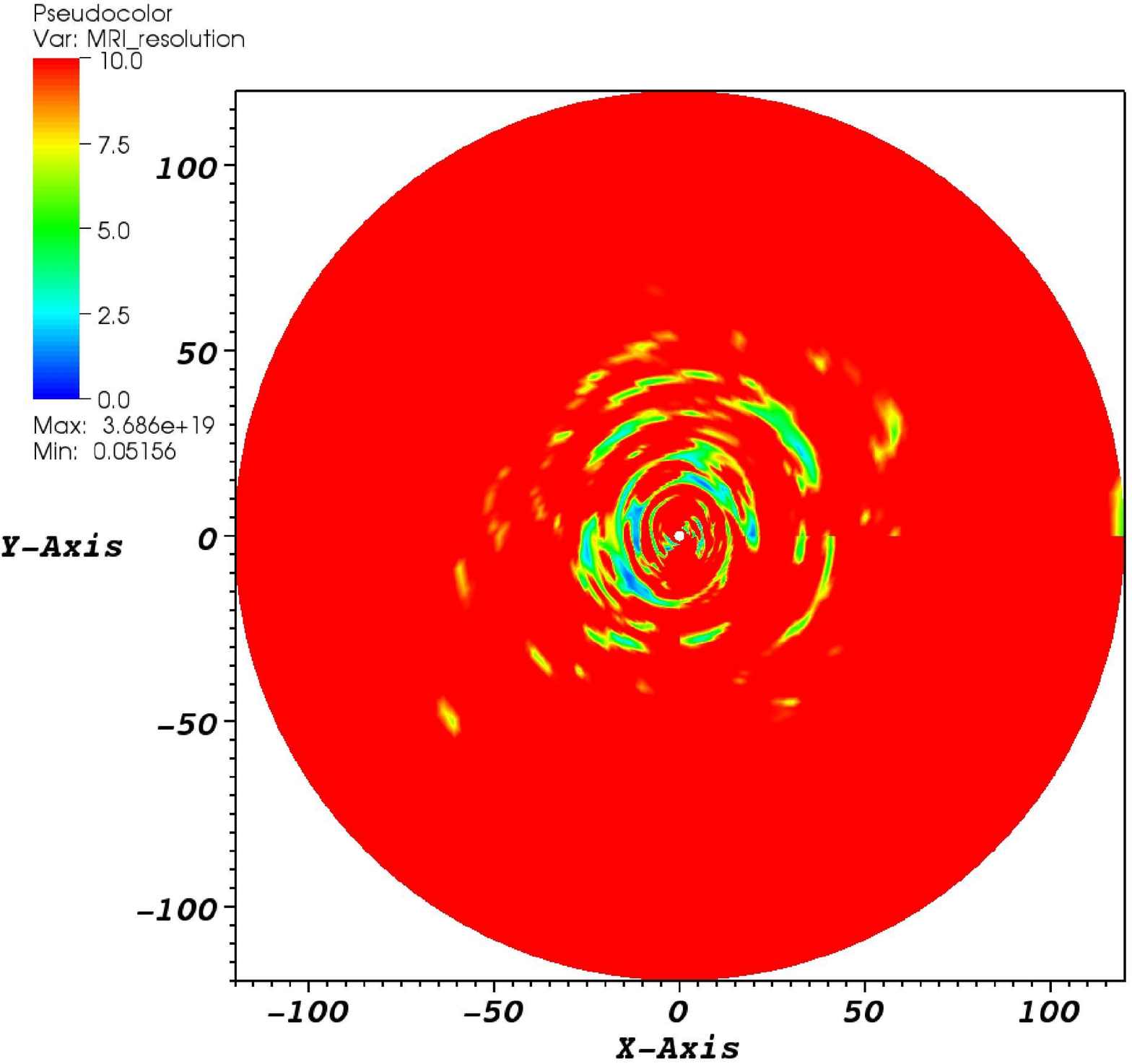}
\caption{
Number of azimuthal grid zones per critical MRI wavelength in the torus' midplane for the untilted
(left) and tilted (right) simulations.  Excepting a few small, isolated, transient patches, the
entire simulation domain has roughly $10$ or more zones per wavelength, indicating MRI turbulence
should be well captured.
}
\label{fig:lambda_mri}
\end{figure*}

Again apart from the tilt, these simulations run on essentially the same nested grids with
equivalent peak resolution of $128^3$ zones realized throughout the bulk of the simulation
domain. These grids have sufficient spatial resolution at all radii to reasonably capture the MRI.
As discussed in \cite{sto96}, several grid zones are required per critical MRI wavelength
($\lambda_{\rm MRI}\equiv 2\pi v_{\rm A}/\Omega=2\pi\sqrt{B^2/4\pi\rho\Omega^2}$,
where $v_{\rm A}$ is the Alfv\'{e}n velocity, $B^2$ is the square of the magnetic field, $\rho$ is
the fluid mass density, and $\Omega$ is the local orbital frequency) to ensure a convergent
treatment of the instability and associated turbulence. As illustrated in figure
\ref{fig:lambda_mri}, there are at least $10$ grid zones per critical wavelength, apart from small,
isolated, transient patches where the field is weak.

We restrict our analysis to a poloidal sector of grid zones within $35^\circ$ of the original outer
torus midplane at all radii and azimuthal angles. Within this region, we uniformly sample fluid
variables at each grid zone $200$ times per orbit (that is, $7$ times per orbital period of the
direct ISCO on the black hole equatorial plane). Our timing analysis begins four orbital periods
after the start of each simulation to ensure fully developed accretion and MRI turbulence, then
extends for six additional orbits.

We study the time variation of three fluid variables, namely, the rest mass density
$\rho$ as measured in the fluid's rest frame and the radial and poloidal components of the
3-velocity: $v^r=u^r/u^t$ and $v^\theta=u^\theta/u^t$. Here, $u^\mu$ are the general relativistic
4-velocity components expressed in tilted Kerr-Schild coordinates \citep{fra07}, measured at zone
centers.

We search for quasi-periodic activity in these variables through analysis of radially dependent
temporal power spectra. The three variables are measured in each grid zone at regular time
intervals, and linear secular changes are subtracted from the zone's time series as in
\cite{sch06}. Other authors have used more elaborate and physically-motivated fitting functions to
accomplish this ``pre-whitening.'' For example, \cite{rey08} advocate exponential forms because of
the secular spreading of mass within the disk which, in our simulation and depending on the
location of a given cell with respect to the initial torus, can act to increase or decrease
density. On the other hand, in a tilted configuration, another significant secular change is
Lense-Thirring precession with a period of roughly $90$ orbits. In the six analyzed orbits, the
disk's angular momentum vector precesses roughly $24^\circ$ about the black hole's spin axis. This
precession induces some fraction of the full period of a sinusoidal modulation at the precession
frequency. Whether this results in an increase or decrease in density is determined by the location
of the given zone with respect to the initial torus. Since these two effects are completely
independent of each other and can act to either increase or decrease the local density, it is
possible for the two to act in opposition. We employ, therefore, a simpler linear (i.e.
model-independent) fit to remove secular changes. Finally, these pre-whitened time series are
windowed using a Bartlett window, zero padded and fast Fourier transformed to produce the zones'
raw spectra \citep{pre92}.

\begin{figure*}
\centering
\epsscale{1.0}
\includegraphics[width=0.48\textwidth]{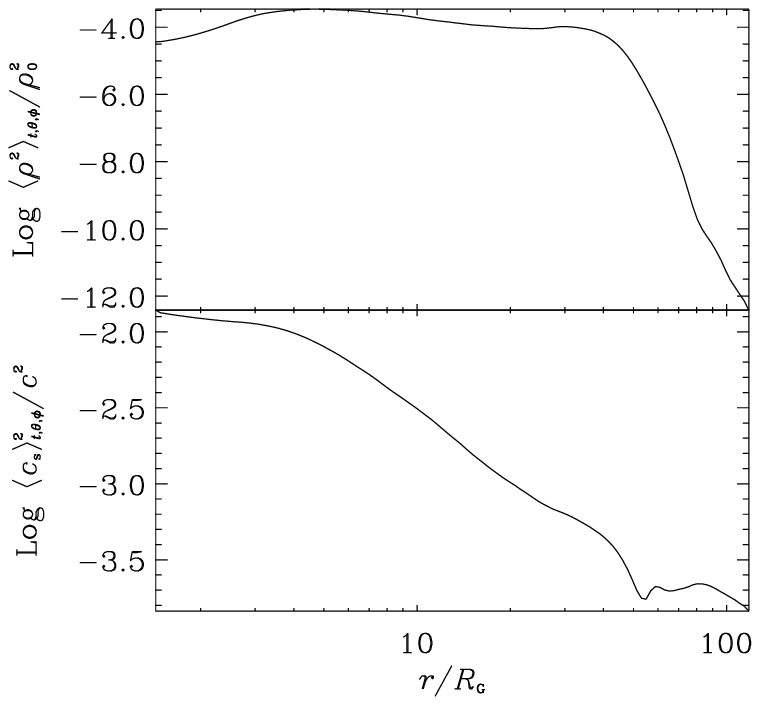}
\includegraphics[width=0.48\textwidth]{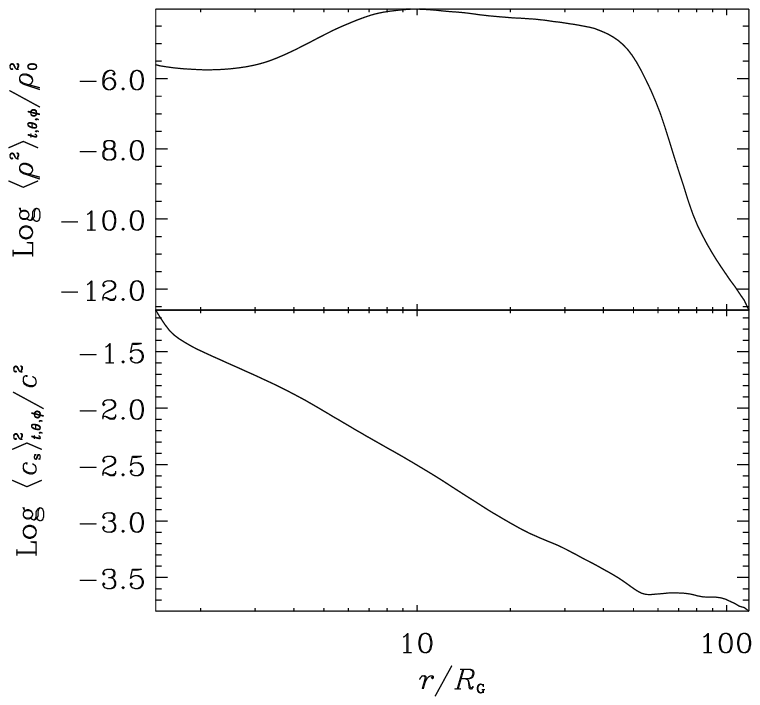}
\caption{
Normalization curves as a function of coordinate radius $r$ showing the mean squared density (top)
and squared mean sound speed (bottom) for both the untilted (left) and tilted (right) simulations,
in simulation units (i.e. the natural density $\rho_0=c^6/G^3M^2$ and the speed of light $c$).
}
\label{fig:norm_r}
\end{figure*}

In calculating the final radially dependent, temporal power spectrum, we average our spectra on shells
of constant coordinate radius and divide by appropriate background scaling factors to enhance the
radial contrast between features using physically relevant scales.  Symbolically,
\begin{eqnarray}
P_\rho(r,f)&\equiv&\langle|\tilde{\rho}|^2\rangle_{\theta,\phi}/\langle\rho^2\rangle_{t,\theta,\phi} \nonumber \\
P_{v_r}(r,f)&\equiv&\langle|\tilde{v_r}|^2\rangle_{\theta,\phi}/\langle c_{\rm s}\rangle_{t,\theta,\phi}^2 \label{eq:psd_norm} \\
P_{v_\theta}(r,f) &\equiv&\langle|\tilde{v_\theta}|^2\rangle_{\theta,\phi}/\langle c_{\rm s}\rangle_{t,\theta,\phi}^2, \nonumber
\end{eqnarray}
where $\tilde{\rho}$, $\tilde{v_r}$, and $\tilde{v_\theta}$ are the pre-whitened, windowed, padded,
and Fourier transformed fluid variables as functions of space and frequency. The weight factors
$\langle\rho^2\rangle_{t,\theta,\phi}$ and $\langle c_{\rm s}\rangle_{t,\theta,\phi}^2$ are the
mean square density and squared average sound speed, $c_{\rm s}\equiv\sqrt{P/\rho}$, respectively.
The radial profiles of these weight factors are shown in figure \ref{fig:norm_r}. We choose the
sound speed as an appropriate scaling speed for both acoustic and inertial modes because both
mode types involve fluid oscillations with vertical structure of order the disk scale
height, $H$, and frequencies comparable to the orbital frequency, $\Omega$. Thus, the relevant
velocity scale is $H\Omega$ which, by vertical hydrostatic equilibrium, is approximately equal to
the sound speed.

\section{Results}

\begin{figure*}
\centering
\epsscale{1.0}
\includegraphics[width=0.48\textwidth]{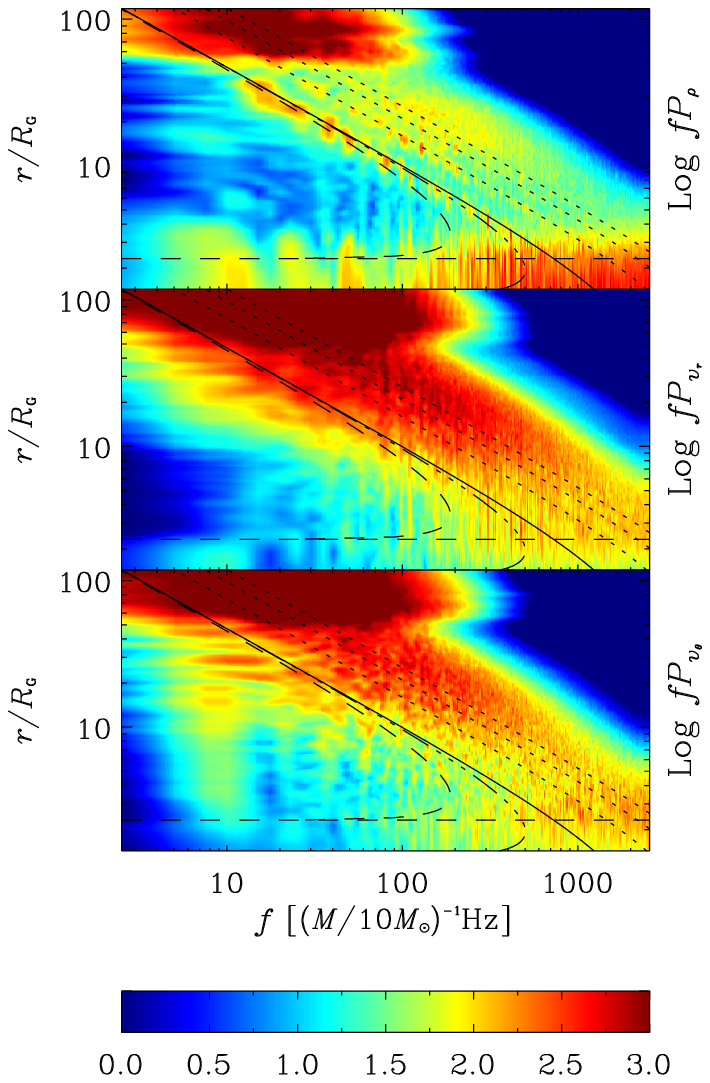}
\includegraphics[width=0.48\textwidth]{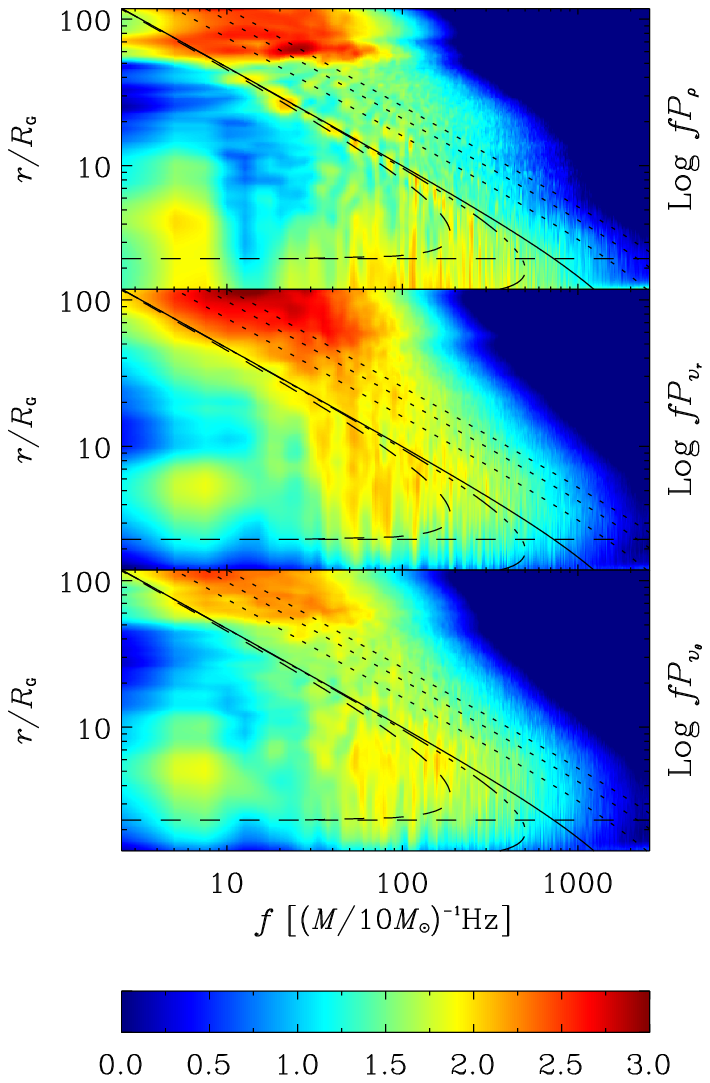}
\caption{
Shell averaged power spectral density as a function of frequency $f$ and coordinate radius $r$ in
density (top), radial velocity component $v_r$ (middle), and polar velocity component $v_\theta$
(bottom), for the untilted (left) and tilted (right) simulations. Overplots include the orbital
frequency (solid) and its harmonics (dotted), the geodesic radial epicyclic frequency and the ISCO
radius (dashed), and the geodesic vertical epicyclic frequency (triple-dot dashed). Note that, for
simplicity, an arbitrary constant has been subtracted from each logarithmic plot, setting the
colorbars' centers at $1.5$.
}
\label{fig:fPu_rf}
\end{figure*}

Figure \ref{fig:fPu_rf} depicts the radial power spectrum in the untilted and tilted simulations.
In the untilted simulation, three spectral features are prominent. First, high power at large
radii in all variables reflects artificially enhanced variability by the extremely low background
density and sound speed weight factors in those regions (see figure \ref{fig:norm_r}). Very little
material is present in these regions since the simulated time period is short when compared with
the diffusion time required to spread the disk to such large radii.

Second, density variability at high frequencies near and within the ISCO is present. We do not
fully understand this behavior. Similar variability seems absent in this simulation's velocity
variables and, as will be discussed later, in all three variables in the tilted simulation. This
suggests that the feature is physical (i.e. it is not a relic of the Fourier transform procedure)
and reflects a fundamental difference in the accretion flow between the two disk configurations.
For a detailed comparison of the titled and untilted flow patterns, see \cite{fra07}.

Third, significant power tracks the orbital frequency (or, arguably, the vertical epicyclic
frequency) in the density spectrum. \cite{rey08} suggest that this power is a result of local
vertical epicyclic oscillations and would predict higher order acoustic oscillations with
frequencies
\begin{eqnarray}
\omega\simeq[n(n\gamma-n+3-\gamma)/2]^{1/2}\kappa_\theta, \label{eq:vert_harm}
\end{eqnarray}
where $n=1,2,3,\ldots$ is a
vertical quantum number, $\gamma$ is the adiabatic index (which is $5/3$ in our simulations), and
$\kappa_\theta$ is the vertical epicyclic frequency \citep{bla06}.\footnote{Note that
$\kappa_\theta\ne\Omega$ for our Kerr spacetime, in contrast to the pseudo-Newtonian spacetime used
by \citet{rey08}. Their simulations did, in fact, exhibit modes with frequencies
$\omega\simeq(n\gamma+1)^{1/2}\kappa_\theta$ with $n=0,1,2,\ldots$, because they employed an
isothermal vertical profile with an adiabatic equation of state. Our simulations are better
described by an adiabatic profile and equation of state, thus acoustic waves should obey the
dispersion relation \ref{eq:vert_harm}.}

While such vertical acoustic modes aptly explain the features seen in \citeapos{rey08}
simulations, they do not appear to correctly describe our spectra. Even though we cannot claim the
power more closely tracks the orbital frequency than the vertical epicyclic frequency, it appears
that the higher order diagonal tracks tightly follow harmonics of the orbital frequency with
integer coefficients (indicated by dotted lines in figure \ref{fig:fPu_rf}) rather than those
curves described by equation \ref{eq:vert_harm} relating to the vertical epicyclic frequency. In
particular, we note this agreement even at frequencies higher than the vertical epicyclic frequency
maximum. Additionally, tracks do not appear strongly in the polar velocity spectrum as would be
expected if acoustic motion in that direction were present. Hence, we argue that the
fundamental dynamics giving rise to our tracks is not related to vertical acoustic motion.

An alternative interpretation for the power along the orbital frequency and its harmonics in our
simulations is that they represent non-axisymmetric clumps orbiting the central black hole on
Keplerian trajectories. One would not expect the presence of such orbiting clumps in the
\cite{rey08} simulations \citep{rey09}. They employed a $30^\circ$ azimuthal wedge with periodic
boundary conditions in $\phi$, so their lowest order, non-axisymmetric structure would, in our
language, have at least an azimuthal quantum number $m=12$. These high $m$ modes would manifest
as fluid oscillations at a frequency $12\Omega$ but would likely be dominated by turbulence in
both \citeapos{rey08} simulations and our own. In addition, it could be that the low $m$ clumps
that we see in our simulations dominate the vertical acoustic waves that they saw in their wedge.

\begin{figure*}
\centering
\epsscale{1.0}
\includegraphics[width=0.48\textwidth]{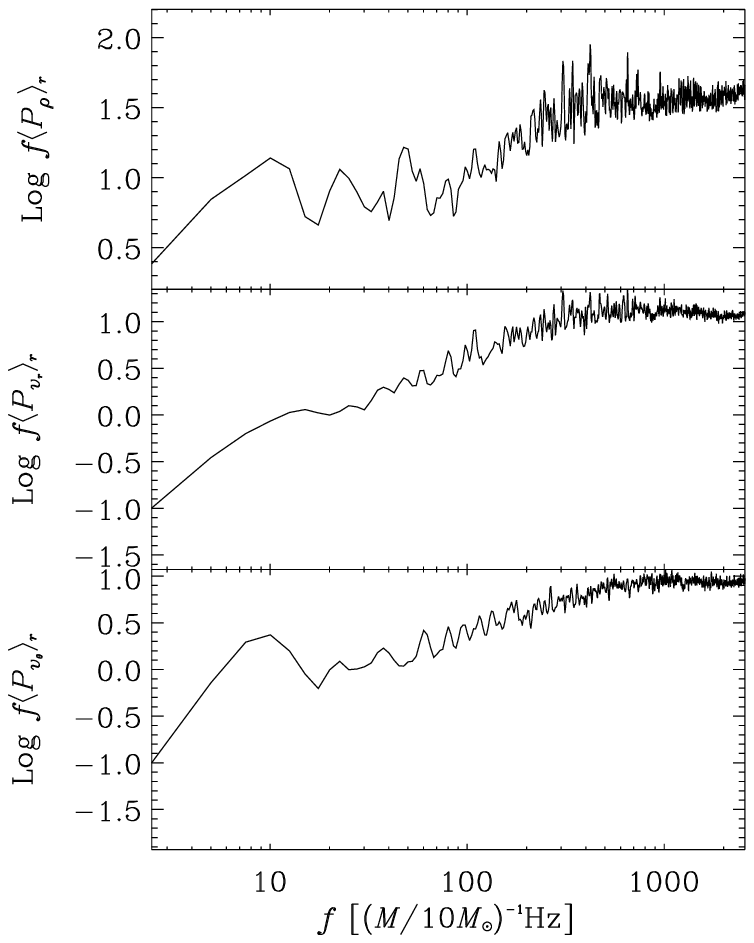}
\includegraphics[width=0.48\textwidth]{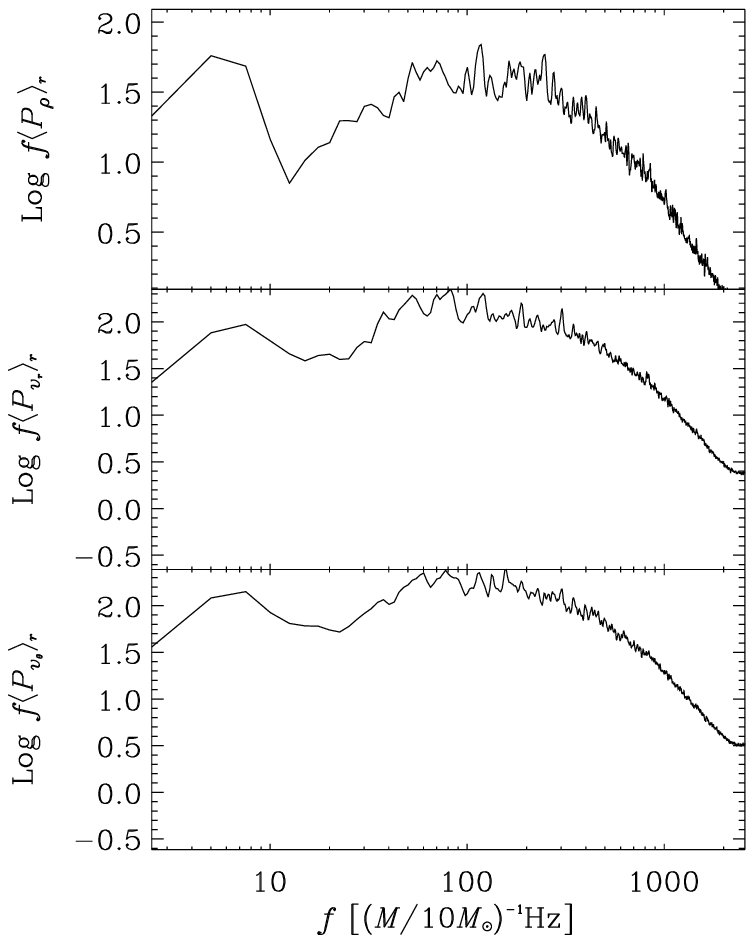}
\caption{
Untilted (left) and tilted (right) disk's power spectra averaged radially between
$R_{\rm ISCO}$ and $10R_{\rm G}$ and on shells as a function of frequency $f$ in density $\rho$
(top), radial velocity $v_r$ (middle), and polar velocity $v_\theta$ (bottom). The curves'
fundamentally different shapes indicate that while the untilted simulation is dominated by
acoustic-like variability, the tilted simulation is better characterized by inertial-like spectral
power.
}
\label{fig:fPu_if}
\end{figure*}

The tilted simulation's power spectrum, shown in figure \ref{fig:fPu_rf}, similarly exhibits power
along the orbital frequency and its harmonics and spurious variability at large radii. In contrast
to the untilted configuration's spectra, no high frequency variability appears near or within the
ISCO in any variable. A characteristically different shape to the overall spectra is
evident: while the untilted spectra are dominated by acoustic-like variability (i.e.
variability at frequencies larger than the orbital frequency), the tilted spectra exhibit
considerably more power at inertial-like frequencies (i.e. frequencies smaller than the orbital
frequency, particularly those bounded above by the radial epicyclic frequency). To visualize the
stark contrast between the the untilted and tilted power spectra, figure \ref{fig:fPu_if} displays
the power integrated in radius between $R_{\rm ISCO}$ ($\sim2.4R_{\rm G}$) and $10R_{\rm G}$.
It is clear that the tilted simulation is contrastingly dominated by power at frequencies
characteristic of inertial variability, whereas the untilted simulation is not. We believe this
power has not been previously observed in MHD simulations.

Figures \ref{fig:fPu_rf} and \ref{fig:fPu_if} may also indicate higher power peaks amid the
continuum of inertial-like variability. Localized in frequency and extended in radius, such peaks
suggest coherent variability arguably spanning the region from the spacetime's ISCO to the radius
at which the orbital frequency equals the oscillation frequency. Additionally, power seems to leak
inwards of the ISCO, particularly in the density spectrum. We can identify matching peaks in each
of the three fluid variables at corresponding frequencies, suggesting that the features are
physical in nature and are not numerical artifacts. Among these, power at $118{\rm Hz}$ is elevated
above the continuum in the density and radial velocity spectra and exhibits a minimal though
identifiable excess in the polar velocity spectrum. We hope to characterize the nature of the
general power at inertial frequencies by carefully analyzing the three dimensional structure of the
variability at $118{\rm Hz}$.

Isothermal diskoseismic models of untilted, flat disks indicate that perturbations with local
radial wave number $k$ obey the dispersion relation \citep{oka87},
\begin{eqnarray}
k^2=\frac{(\hat{\omega}^2-\kappa_r^2)(\hat{\omega}^2-n\kappa_\theta^2)}{\hat{\omega}^2c_{\rm s}^2},\label{eq:disp_rel}
\end{eqnarray}
where $\hat{\omega}\equiv\omega-m\Omega$ is the Doppler-shifted wave frequency and $\kappa_r$ is
the radial epicyclic frequency.\footnote{\cite{oka87} assume an isothermal equation of state and
vertical profile. For an adiabatic equation of state and vertical profile, the acoustic factor must
be slightly modified as in equation \ref{eq:vert_harm} above. However, for the important $n=0$ and
$n=1$ modes, both dispersion relations agree. Physically, these two modes exhibit no vertical oscillation
and a bulk vertical oscillation of the local center of mass, respectively. There is therefore no
internal compression, and the thermodynamics are irrelevant.}
Wave propagation is only possible where $k^2>0$ and equation \ref{eq:disp_rel} is quadratic
in $\hat{\omega}^2$, so it is clear that two distinct oscillation modes exist for a given frequency
$\omega$ and set of vertical and azimuthal wave numbers $n=0,1,2\ldots$ and $m=0,1,2\ldots$,
respectively.  As summarized in \cite{fer08}, modes that have pressure as their dominant restoring
force ($p$-modes) can exist at radii where
$\omega<m\Omega-{\rm max}(\kappa_r, \sqrt{n}\kappa_\theta)$ and
$\omega>m\Omega+{\rm max}(\kappa_r, \sqrt{n}\kappa_\theta)$, while modes whose dominant restoring
force is inertial in character ($r$-modes) can lie where
$m\Omega-{\rm min}(\kappa_r, \sqrt{n}\kappa_\theta)<\omega<m\Omega+{\rm min}(\kappa_r, \sqrt{n}\kappa_\theta)$.
Outside a given mode's trapping region (i.e. where $k$ is imaginary), the oscillation damps
exponentially.
Unfortunately, this diskoseismic theory applies to perturbations of a geometrically thin, flat
accretion disk.  It is therefore difficult to compare its predictions directly to a pressure supported,
moderately tilted disk due to the nonlinear effects of spatially varying tilt and twist and
significant vertical thickness.

Power at $118{\rm Hz}$ displays complex radial structure. Two local maxima appear just inside the
inner and outer edges of the radial epicyclic frequency, and another lies just within the orbital
frequency's corotation radius. The first two features may imply a superposition of modes trapped
within respective resonances, while the third has the characteristics of the aforementioned
Keplerian clumps.  There is also significant mode power well inside the equatorial test particle
ISCO, which might be due to several factors.  Pressure gradients can sustain stable fluid element
orbits down to a smaller effective ISCO radius. Additionally, mode power may manifest itself inside
this effective ISCO due to active accretion; any mode with a trapping radius close to the plunging
region may regularly slough off material to smaller radii at the mode frequency. Such fluid pulses
would reveal themselves in spectra as higher power at all radii within the plunging region.

\begin{figure*}
\centering
\epsscale{1.0}
\includegraphics[width=0.98\textwidth]{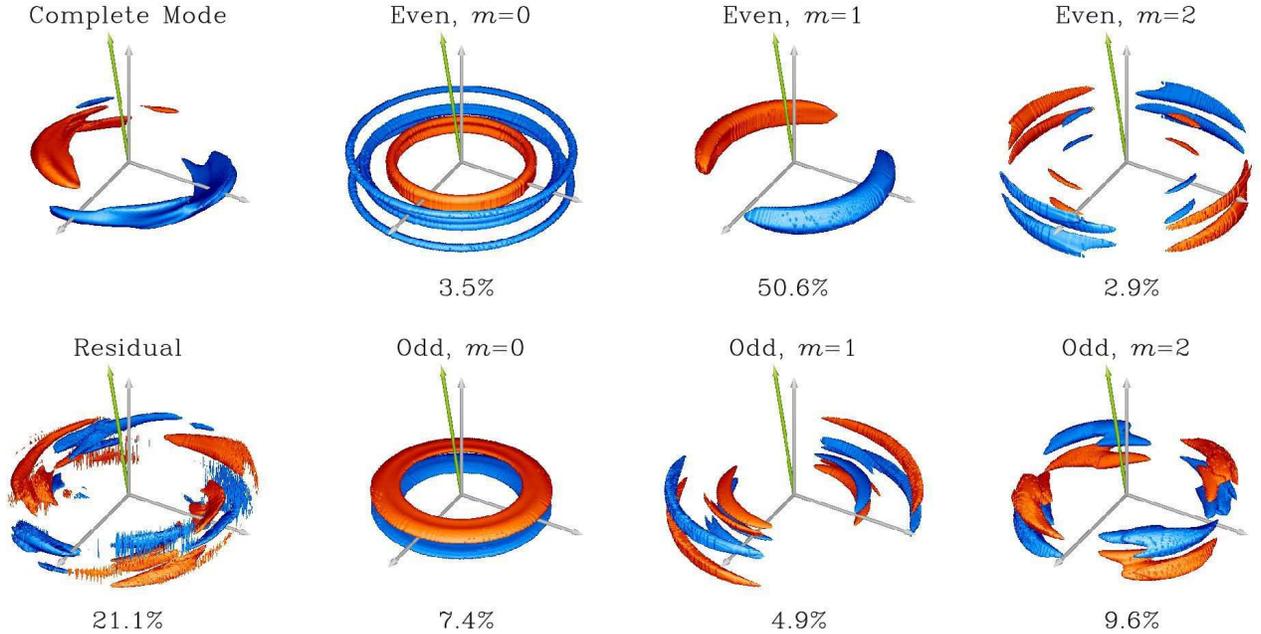}
\caption{
Positive (red) and negative (blue) isosurfaces of density fluctuations evaluated at $118{\rm Hz}$
in the tilted disk simulation.
Subfigures include the unmodified complete mode structure (leftmost top), projections onto six
orthogonal basis functions having even or odd parity in the polar direction and azimuthal quantum
numbers $m=0,1,$ or $2$ (right three columns), and the residual, that is, the difference
between the complete mode and the sum of the six projections (leftmost bottom). The gray axes
represent a rotated coordinate system in which the $z$-axis is normal to the average orientation of
the disk midplane. Parity is defined with respect to this $xy$-plane. A green vector indicates the
black hole spin axis.
Also given is the fraction of the total mode power in each projection. The surface levels are drawn
at half the maximum of the unmodified mode's absolute value times the square root of the respective
projection's fractional power (or simply times $1$ in the unmodified case).
While this static plot depicts the real (or zero phase) component of the
eigenmode, the online animated figure varies the mode's complex phase as a function of time.
}
\label{fig:eigen3d}
\end{figure*}

To quantify the radial structure and better understand the physical nature of the $118{\rm Hz}$
variability, we extract its spatial structure from the real and imaginary parts of the Fourier
transform of our fluid variables. Diskoseismic eigenmodes possess well defined azimuthal wave
numbers $m$ and parity with respect to the disk equatorial plane. Because a tilted disk lacks both
axisymmetry and vertical reflection symmetry, its wave eigenmodes need not be characterized by such
quantum numbers.  Still, in order to connect to the analytic theory of untilted disks, we project
the $118{\rm Hz}$ density structure onto six orthogonal basis functions with well-defined $m$
and parity. Figure \ref{fig:eigen3d} depicts 3D isosurfaces of the real part of the eigenfunction
and its projections; an animation of these isosurfaces may be found online with this publication. A
clear $m=1$ structure with even parity is the dominant component, accounting for roughly $50\%$ of the
total power. The next two most powerful components are the $m=0$ and $m=2$ modes with odd parity,
possessing roughly $10\%$ and $7\%$ of the total power, respectively.

\begin{figure}[b]
\epsscale{1.0}
\includegraphics[width=0.48\textwidth]{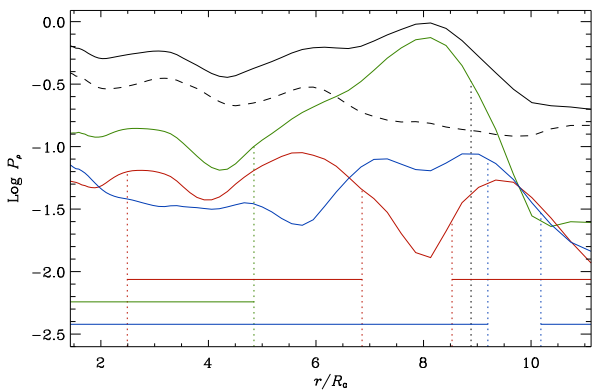}
\caption{
Radial distribution of power at $118{\rm Hz}$ (black), its (even, $m=1$)
component (green), (odd, $m=0$) component (red), (odd, $m=2$) component (blue), and residual
(dashed black). Colored, vertical, and dotted lines indicate the zeros of the squared radial wave
number $k^2$ for each mode, where even parity has been naively associated with $n=0$ and odd with
$n=1$. Horizontal lines designate the regions in which $k^2$ is positive, that is, the regions
permitting wave propagation according to equation \ref{eq:disp_rel}. The orbital frequency is
indicated by a black, vertical, dotted line.
}
\label{fig:eigen3d_power_r}
\end{figure}

The radial distribution of power in each of these projected modes sheds light on the physical
nature of the oscillations. Figure \ref{fig:eigen3d_power_r} illustrates the total $118{\rm Hz}$
power and each of its three most dominant mode components as a function of radius.
Additionally, vertical dotted lines indicate the zeros of $k^2$ for each mode's quantum numbers
and horizontal lines show the regions in which $k^2$ is positive, that is, the mode propagation
regions. For the (odd, $m=1$) and (odd, $m=2$) modes, local maxima occur within predicted trapping
regions, suggesting that where the $118{\rm Hz}$ feature is most significant, the former behaves as
an $r$-mode bound by the radial epicyclic frequency, and the latter is a $p$-mode trapped between 
$2\Omega-\kappa_\theta$ and a reflecting boundary near the plunging region.  Conversely, the
(even, $m=1$) mode peaks near the orbital frequency (indicated by the vertical, black dotted line)
and well inside a region forbidden by the diskoseismic models. Recalling the trail of high power
peaks tracking the orbital frequency in figure \ref{fig:fPu_rf}, we speculate that this (even,
$m=1$) component is an orbiting clump and not a trapped inertial or acoustic mode.

However, we note an alternative explanation for power in the (odd, $m=0$) and (odd, $m=2$)
components. Consider the following: suppose we have a pure (even, $m=1$) mode but that its
rotation axis is slightly misaligned with the $z$-axis of a cylindrical coordinate system. If we
decompose this configuration with respect to the true $z$-axis, we will find some small amount of
power has spilled into the other modes, most notably into the (odd, $m=0$) and (odd, $m=2$) modes.
In a practice, it is difficult to accurately determine the position of the disk midplane, and it is
possible our decomposition mixes power between modes. However, the correlations between the
predicted mode trapping regions and the radial power profile suggest the power distribution is not
entirely a numerical artifact.

In addition to the nonlinear excitation of trapped waves by warps and eccentric orbits within the disk
flow, a titled accretion flow may be more conducive to mode trapping because of the distinct character
of its plunging region.  In contrast with an untilted disk which has a largely axisymmetric plunging
region, mass accretion at small radii in our tilted simulation is confined to two discrete plunging
streams \citep{fra07}.
Thus, it is conceivable that the azimuthal sections of the inner disk away from these streams
provide significant radial reflection of traveling waves in the disk.

\section{Conclusions}

Previous work has demonstrated that inertial modes are vulnerable to disruption by MRI turbulence
\citep{arr06,rey08}. Our Fourier analysis of the untilted simulation, a configuration devoid of
tilt or eccentricity, corroborates this conclusion. However, nonlinear analytic calculations
\citep{kat04a, kat08, fer08} have indicated that disk warps and eccentricity may excite inertial
modes; both of these features are generically present in tilted accretion flows. Though the
analysis of the variability in our tilted simulation at $118{\rm Hz}$ does not appear to be wholly
inertial (or even acoustic) in nature, the presence of two weak, odd parity, $m=0$ and $m=2$
components provides some support for this hypothesis.

Evidence that disk tilt is connected with the excitation of inertial or acoustic waves is
twofold. First, fundamentally different shapes in the spectra from the untilted and tilted
simulations implies a correlation between tilt and the power at frequencies characteristic of
inertial waves. Second, the radial structure of the variability suggests a superposition of
modes, at least one of which may be described as inertial in nature in the context of relevant
analytic models. Together, these clues establish the presence of variability unique to a tilted
geometry which appears to be at least partially inertial in character.

Recalling that a tilted disk configuration naturally yields a low frequency QPO activity in black
hole X-ray binaries and speculating that our work's inertial-like variability may be related to
high frequency QPO activity in the same systems, we advocate further numerical exploration of these
tilted geometries. Torques associated with the Kerr spacetime can produce a bodily precession of the
tilted disk at frequencies within the $0.1$ and $30{\rm Hz}$ range associated with the low frequency
QPO. Because this precession frequency is strongly dependent on the radial extent of the disk in
this model, it readily explains the observed correlation between the LFQPO's frequency
and disk flux \citep{rem06,fra08,ing09}. Similarly, our current work shows inertial or acoustic
variability excited and maintained by the disk tilt that exhibits frequencies of order the radial
epicyclic frequency maximum. Such frequencies fall within the observed range of $40$ to
$450{\rm Hz}$ for HFQPOs in stellar mass black hole systems.


HFQPOs are only observed when a system's X-ray emission exhibits the characteristics of the steep
power-law state (cf. \citealt{rem06} for a review of the properties of X-ray black hole binaries),
and we must acknowledge, however, that our simulations are almost certainly not accurate
representations of that state.  For instance, since they neither account for radiative cooling nor
capture all forms of dissipation, our simulations likely do not reproduce some of the key features
found in various models for the state's structure (e.g. the energetically significant corona in
\citealt{don07}). However, we are confident that our simulations and analysis provide insight into
the dynamical effects of a tilted accretion flow, particularly, the excitation of inertial and
acoustic modes.

\sloppypar{
We thank Chris Done, Paul Henisey, Mami Machida, Phil Marshall, Gordon Ogilvie, Chris Reynolds, and
Alexander Tchekhovskoy for useful conversations, and Sam Cook for his help with data transport and
preparation. We also thank the anonymous referee for comments that significantly improved this
paper.
This work was supported in part by NSF grant AST-0707624.  The work of BTF was supported by FCT
(Portugal) through grant SFRH/BD/22251/2005. PCF acknowledges support from NSF grant AST-0807385
and an American Astronomical Society International Travel Grant.
We are also grateful to the Nordic Institute for Theoretical Astrophysics for
hosting a workshop on QPOs where much of this work was completed.
}

\end{document}